\documentclass{kapproc} 
\usepackage{procps}
\usepackage[dvips]{graphicx}
\upperandlowercase
\kluwerbib

\def\simlt{\mathrel{\rlap{\lower 3pt\hbox{$\sim$}} \raise
        2.0pt\hbox{$<$}}} \def\simgt{\mathrel{\rlap{\lower
        3pt\hbox{$\sim$}} \raise 2.0pt\hbox{$>$}}}
\def\spose#1{\hbox to 0pt{#1\hss}}
\def\lta{\mathrel{\spose{\lower 3pt\hbox{$\mathchar"218$}}
     \raise 2.0pt\hbox{$\mathchar"13C$}}}
\def\gta{\mathrel{\spose{\lower 3pt\hbox{$\mathchar"218$}}
     \raise 2.0pt\hbox{$\mathchar"13E$}}}

\newcommand{\msun}{\,{\rm M_\odot}}

\begin{document}

\articletitle[The Merging History of Massive Black Holes]
{The Merging History of Massive Black Holes}

\author{Marta Volonteri,\altaffilmark{1} Francesco Haardt,\altaffilmark{2}
Piero Madau\altaffilmark{1} \& Alberto Sesana\altaffilmark{2}}

\altaffiltext{1}{University of California, Santa Cruz, CA 95064, USA }
\altaffiltext{2}{Universit\'a dell'Insubria, Como, Italy}

\anxx{Volonteri\, Marta}

\begin{abstract}
We investigate a hierarchical structure formation scenario describing the
evolution of a Super Massive Black Holes (SMBHs) population. The seeds of
the local SMBHs are assumed to be 'pregalactic' black holes, remnants of
the first POPIII stars. As these pregalactic holes become incorporated
through a series of mergers into larger and larger halos, they sink to the
center owing to dynamical friction, accrete  a fraction of the gas in
the merger remnant to become supermassive, form a binary system, and
eventually coalesce. A simple model in which the damage done to a stellar cusps by
decaying BH pairs is cumulative is able to reproduce
the observed scaling relation between galaxy luminosity and core size.
An accretion model connecting quasar activity with major mergers and
the observed BH mass-velocity dispersion correlation reproduces
remarkably well the observed luminosity function of
optically-selected quasars in the redshift range 1<z<5. 
We finally asses the potential observability of the 
gravitational wave background generated by the cosmic evolution of SMBH binaries 
by the planned space-born interferometer LISA. 
\end{abstract}

\begin{keywords}
cosmology: theory -- black holes -- galaxies: evolution -- 
quasars: general
\end{keywords}

\section{Introduction}
In \cite{paperI} (Paper I) we have assessed a model 
for the assembly of SMBHS at the center of galaxies that trace their hierarchical 
build-up far up in the dark halo `merger tree'.  We have assumed that the first
`seed' black holes (BHs) had intermediate masses, $m_{seed}\approx 150\, m_\odot$,
and formed in (mini)halos collapsing at $z\sim 20$ from high-$\sigma$ density 
fluctuations (cfr. \cite{mr01}). 
These pregalactic holes evolve in a hierarchical fashion, following the merger 
history  of their host halos.  During a merger event BHs approach each other owing to 
dynamical friction, and form a binary system. Stellar dynamical processes drive the binary to 
harden and eventually coalesce. 

The merger history of dark matter halos 
and associated black holes is followed through Monte Carlo realizations of the merger 
hierarchy (merger trees) which allow to track the evolution of SMBH binaries along 
cosmic time, and analyze how their fate is influenced by the environment (e.g stellar density cusps).

\section{Accretion history}
The first stars must have formed out of metal-free gas, with the lack of an efficient cooling 
mechanism possibly leading to a very top-heavy initial stellar mass function. If stars form above
260 $m_\odot$, after 2 Myr they would collapse to massive BHs
containing at least half of the initial stellar mass. 
The mass density parameter of our `3.5-$\sigma$' pregalactic 
holes is $\Omega_{seed}\geq 2\times 10^{-9}\,h$.
This is much smaller than the density parameter of the supermassive variety found in the 
nuclei of most nearby galaxies, $\Omega_{\rm SMBH}\approx 2\times 10^{-6}$,
Clearly, if SMBHs form out of very rare Pop III BHs, the \emph {present-day mass 
density of SMBHs must have been accumulated 
during cosmic history via gas accretion, with BH-BH mergers playing a secondary
role}. This is increasingly less true, of course, if the seed holes are more numerous 
and populate the 2- or 3-$\sigma$ peaks instead, or halos with smaller masses 
at $z>20$ (\cite{mr01}). 

To avoid introducing additional parameters to our model, as well as 
uncertainties linked to gas cooling, star formation, and supernova feedback, 
we adopt a simple prescription for the mass accreted by a SMBH during each major 
merger assuming that in every accretion episode the BH accretes a mass proportional 
to the observed correlation between stellar velocity dispersion and SMBH mass.
The normalization factor is of order unity and is fixed 
in order to reproduce both the stellar velocity dispersion and SMBH mass relation 
observed locally and the optical LF of quasars in the redshift range 1<z<5.

\section{Dynamical evolution of BH binaries}
During the merger of two halo$+$BH systems of 
comparable masses, dynamical friction drags in the satellite hole toward the 
center of the newly merged system,
leading to the formation of a bound BH binary in the violently relaxed stellar
core (\cite{bbr}). The subsequent evolution of the binary is determined by the initial 
central stellar distribution. As the binary separation 
decays, the effectiveness of dynamical friction slowly declines and the BH pair 
then hardens via three-body interactions, i.e., by capturing and ejecting at 
much higher velocities the stars passing close to the binary (\cite{qui96}).
The hardening of the binary modifies the stellar density profile,
removing mass interior to the binary orbit, 
depleting the galaxy core of stars, and slowing down further hardening. 
If the hardening continues sufficiently far,
gravitational radiation losses finally take over, and the two BHs coalesce 
in less than a Hubble time.
The merger timescale is computed adopting a simple 
semi-analytical scheme that qualitatively reproduces the evolution observed
in N-body simulations. 

If was first proposed by \cite{ebi91} that the heating 
of the surrounding stars by a decaying SMBH pair would create a low-density 
core out of a preexisting cuspy stellar profile. 
The ability of SMBH binaries in shaping the central structure of galaxies
depends also on how galaxy mergers affect the inner stellar density profiles,
i.e. on whether cores are preserved or steep cusps are regenerated during
major mergers. To bracket the uncertainties and explore  
different scenarios we run two different sets of Monte Carlo realizations
(\cite{paperII}, Paper II).
In the first (`{ cusp regeneration}') we assume, that the 
stellar cusp $\propto 
r^{-2}$ is promptly regenerated after every major merger event. 
In the second (`{ core preservation}') the effect
of the hierarchy of SMBH binary interactions is instead cumulative. 
The simple models for core creation
described above yields core radii that scale almost linearly with galaxy mass, 
$r_c\propto M_0^{0.8\div0.9}$ in the range $10^{12}<M_0<4\times10^{13}\,m_\odot$.
A similarly scaling relation was observed by \cite{fab97} in a sample of local galaxies.
A better test of our model predictions against galaxy data is provided
by the `mass deficit', i.e. the mass in stars that must be added to the observed cores 
to produce a stellar $r^{-2}$ cusp (see fig. 1). The cusp regeneration model 
clearly underestimates the mass deficit observed in massive `core' galaxies, while the
core preservation one produces a correlation between `mass 
deficit' and the mass of nuclear SMBHs with a normalization and slope that are
comparable to the observed relation. 

\begin{figure}
\includegraphics[height=4cm, width=5cm]{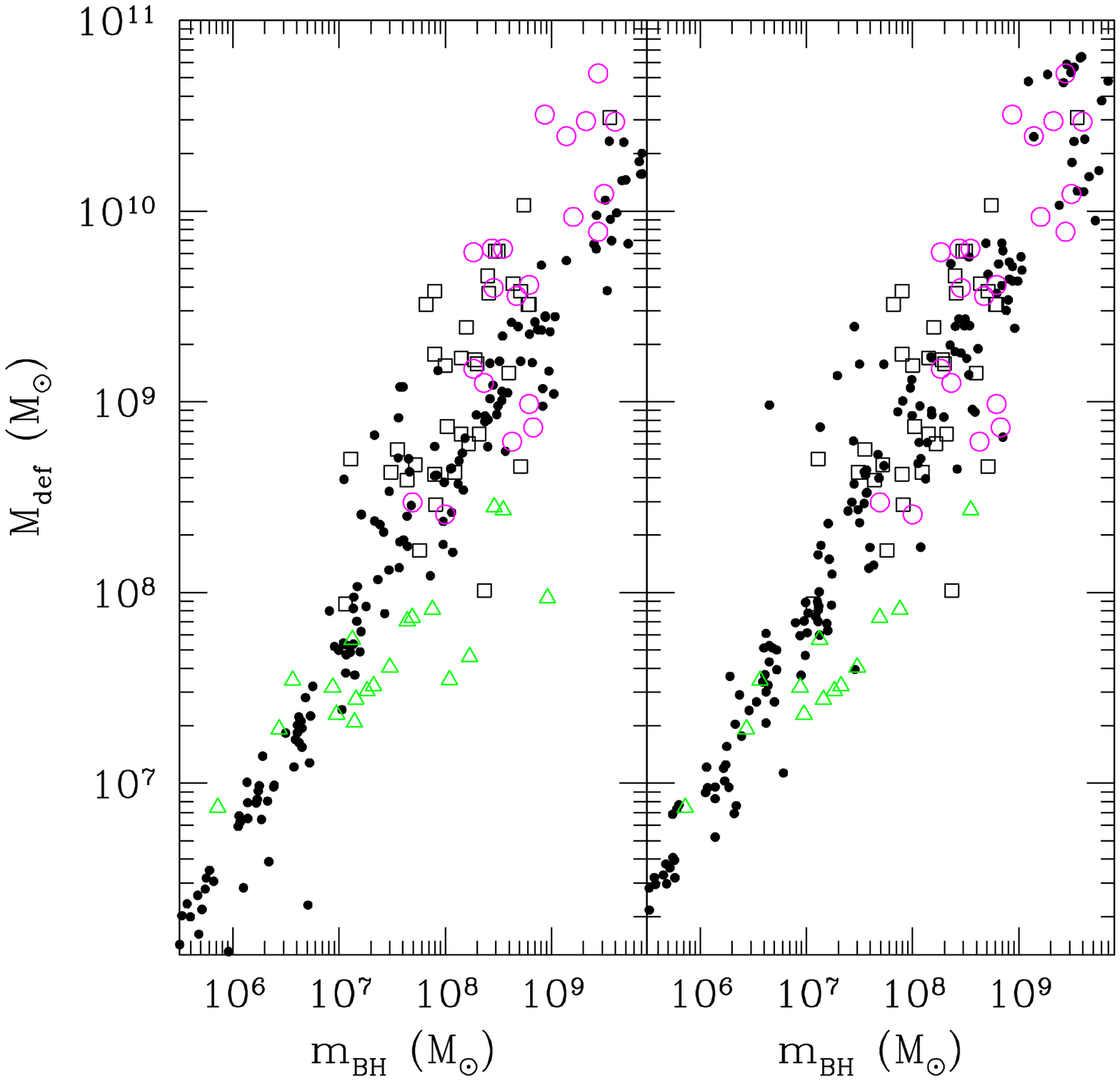}
\includegraphics[height=4cm, width=5cm]{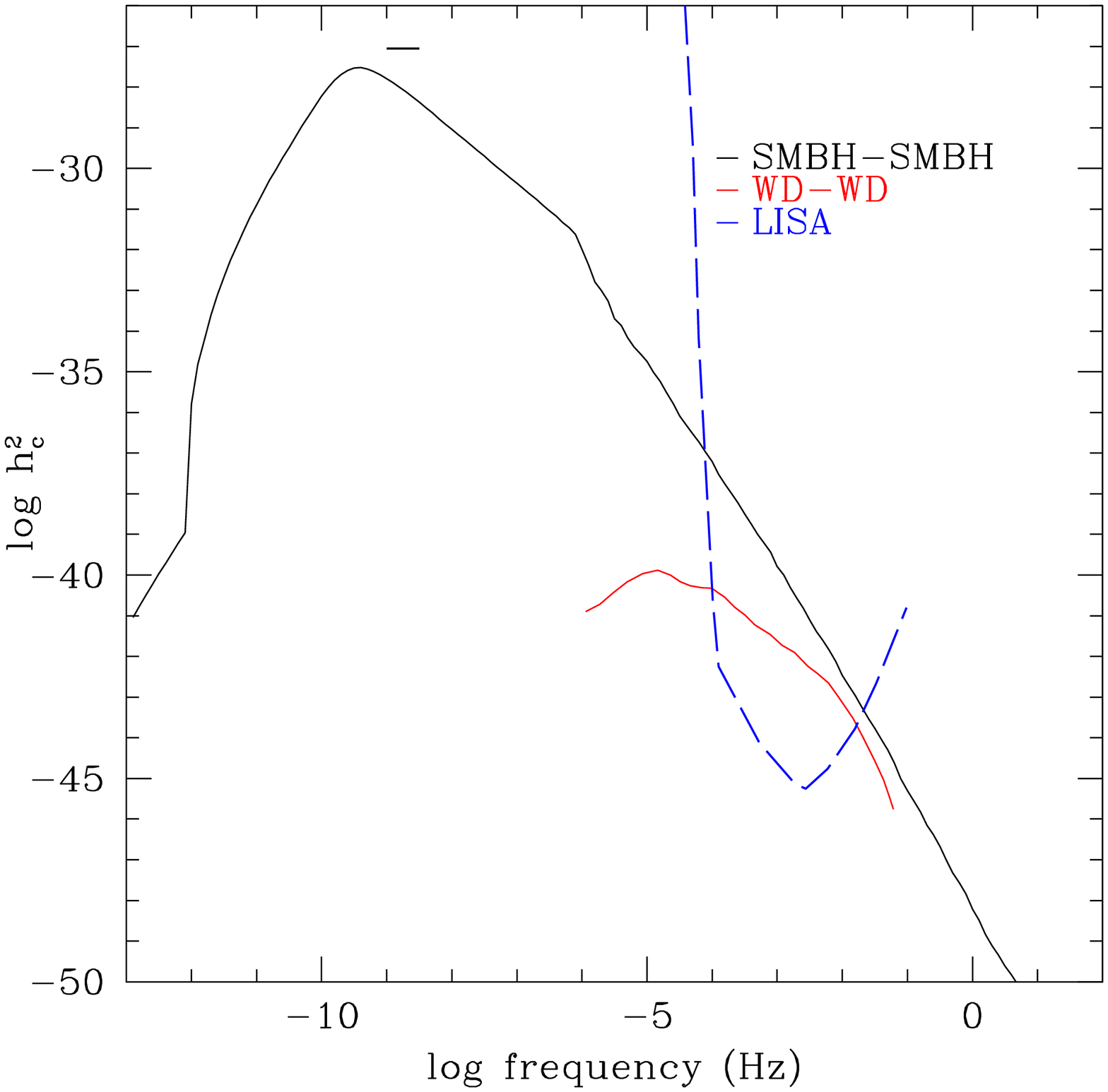}
\sidebyside
{\caption{Mass deficit produced at $z=0$ by shrinking SMBHs 
in our merger tree as a function of nuclear SMBH mass ({ filled dots}). 
{ Left panel:} cusp regeneration case. { Right panel:} core preservation case. 
Galaxies without a core are shown as vertical arrows at an arbitrary mass deficit 
of $10^6m_\odot$. 
{Empty squares, circles and triangles}: mass deficit inferred in a sample of local 
galaxies by \cite{mil02} and \cite{fab97}.}}
{\caption{Gravitational background from the ensemble
of cosmic SMBH binaries. The characteristic strain squared is plotted vs.
the wave frequency (solid line). Dashed line: LISA sensitivity window for one year 
observation. Thick dash at $f\simeq 10^{-9}$: current limits from pulsar
timing experiments (\cite{lommen}).
Lower solid line: the spectrum expected from the extragalactic 
population of white dwarf binaries from \cite{farmer}.}}
\end{figure}

\section{Gravitational waves from SMBH binaries}
We have computed the gravitational wave (GW) background (in terms of the
characteristic strain spectrum) due to the cosmic population of SMBHs evolving along the lines discussed 
in the previous sections. Details and a complete discussion can be found in \cite{gwbpap}.  

Using a linearized GR theory of GW emission and propagation,
framed in a cosmological context, we find that the broad band GW spectrum can be divided into three main
different regimes (fig. 2): for frequencies $\lta 10^{-9}$ Hz, the background 
is shaped by SMBH binaries in the three-body interactions
regime (orbital decay driven by stellar scattering); in the intermediate
band, $10^{-9}\lta f \lta 10^{-6}$ Hz, GW emission itself accounts for most of the potential energy losses,
and the strain has the "standard" $f^{-2/3}$ behaviour; finally, for $f\gta 10^{-6}$, the GW spectrum 
is formed by
the convolution of the emission at the last stable orbit from individual binaries. In the first two regimes,
the strain is dominated by rare low redshift events ($z\lta 2$) involving BHs genuinely supermassive.
At larger frequencies the contribute from smaller and smaller masses starts dominating. 

We have considered the possible future observability of the GW background by the planned
LISA interferometer, showing that one year observation would suffice to detect the GW background 
due to SMBH binaries. Such background should overcome that due to extragalactic WD-WD binaries.
In the LISA
window ($10^{-4}\lta f \lta 0.1$ Hz), the main GW sources are BH binaries
in the mass range $10^{3}\lta M \lta 10^{7}$ $\msun$, with a relevant contribution from $3\lta z \lta 20$.
In the nHz regime, probed by pulsar timing experiments,
the amplitude of the characteristic strain from coalescing SMBH binaries is close to current experimental limits (Lommen 2002, see also \cite{wyitheloeb}).

\chapbblname{vh}
\chapbibliography{vh.bbl}

\end{document}